\documentclass[aps, twocolumn, pra, showpacs, superscriptaddress]{revtex4}
\usepackage{graphicx}
\usepackage{dcolumn}
\usepackage{bm}
\usepackage{amsmath}

\begin{document}

\title{Scaling of quantum Zeno dynamics in thermodynamic systems}
\author{Wing Chi Yu}
\affiliation{Department of Physics and ITP, The Chinese University of Hong Kong, Hong
Kong, China}
\author{Li-Gang Wang}
\affiliation{Department of Physics and ITP, The Chinese University of Hong Kong, Hong
Kong, China}
\affiliation{Department of Physics, Zhejiang University, Hangzhou 310027, China}
\author{Shi-Jian Gu}\email{sjgu@phy.cuhk.edu.hk}
\affiliation{Department of Physics and ITP, The Chinese University of Hong Kong, Hong
Kong, China}
\author{Hai-Qing Lin}
\affiliation{Department of Physics and ITP, The Chinese University of Hong Kong, Hong
Kong, China}
\date{\today}

\begin{abstract}
We study the quantum Zeno effect (QZE) in two many-body systems,
namely the one-dimensional transverse-field Ising model and the
Lipkin-Meshkov-Glick (LMG) model, coupled to a central qubit. Our
result shows that in order to observe QZE in the Ising model, the
frequency of the projective measurement should be of comparable
order to that of the system sizes. The same criterion also holds in
the symmetry broken phase of the LMG model while in the model's
polarized phase, the QZE can be easily observed.
\end{abstract}

\maketitle




\section{Introduction}

Quantum Zeno effect (QZE) refers to the inhabitation of the unitary
time evolution of a quantum system by repeated frequent measurements
\cite {Misra,QZE1,QZE2,QZE3,QZE4,QZE5,QZE6}. It is a phenomena which
is intrinsically related to the projection postulate in quantum
mechanics. For an isolated quantum system, its state vector
undergoes an unitary evolution according to the Schr\"{o}dinger's
equation. At the very beginning, the survival probability for the
system to remain in the initial state changes quadratically with the
elapsed time, while it decreases exponentially in a large time
scale. If a measurement is performed to check whether the system is
still in the initial state or not after some time, the projection
postulate states that the system's state will collapse to the
initial state immediately after the measurement. Therefore, when
successive measurements are performed at a time interval within the
quadratic decaying region, the unitary evolution of the system is
found to be suppressed. In the limit of continuous measurement, the
dynamic evolution of the system is frozen.

The possibility of such an effect to be found in an unstable quantum
system was first pointed out by Fonda \emph{et al.} \cite{1st1} and
Degasperis \emph{et al.} \cite{1st2}. They suggested that the decay
rate of an unstable system may be dependent on the frequency of
intermediate measurements. Later, Misra \emph{et al.}, who gave the
effect its name, presented a general formulism concerning the
semigroups to analyze the QZE \cite{Misra}. Experimentally, the QZE
was first manifested in an radio-frequency transition between two
laser-cooled Beryllium ion ground-state levels \cite{Itano}.

In the recent decade, QZE has been studied intensively within the
content of quantum optics. Among those analyses, the systems under
consideration are only of a few levels in which the QZE can be
easily observed by frequent measurements. However, little attention
of the possibility to observe QZE in quantum many-body systems has
been paid so far. The primary motivation of our work is to
investigate the QZE from the viewpoint of condensed matter physics
and obtain criteria on how frequent the measurements should be
compared to the system size in order to preserve the QZE in
many-body systems.

Mathematically speaking, we are interested in the scaling behavior
of the short time limit of the survival probability of the system's
initial state under the influence of an external perturbation. We
presented a general analysis and find that the leading term of the
survival probability is equal to the fluctuation in the interaction
Hamiltonian. Specifically, we take the one-dimensional
transverse-field Ising model and the Lipkin-Meshkov-Glick (LMG)
model as examples. The leading term of the survival probability in
these two models are calculated analytically and their scaling
behaviors is obtained. In our analysis, we allowed these many-body
models to couple to a central qubit. The significance of the central
qubit is just to provide an external perturbation in the Hamiltonian
and allow us to study how the system's evolution is affected by the
external perturbation in the short time regime. A point to note here
is that the central qubit only serve as a source of external
perturbation and how it evolves is not of our interest. Moreover, we
would like to point out that the leading term of the survival
probability in our study is just the same as the linear response in
the study of quench dynamics \cite{QD1, QD2, QD3}.

The paper is organized as follows: In Section \ref{sec_formulism},
we first present a general formulism for obtaining the survival
probability of the initial state in the short-time limit. Then in
Section \ref{sec_Ising} and \ref{sec_LMG}, we take the
one-dimensional transverse-field Ising model and the
Lipkin-Meshkov-Glick (LMG) model as examples respectively to
illustrate the criteria for observing the QZE, in terms of the size
dependence of the leading term of the survival probability in the
short-time limit. Our analysis shows that in order to observe the
QZE, the number of measurements have to be comparable to the size of
the system. Finally, a summary would be given in Section
\ref{sec_sum}.

\section{Formulism}
\label{sec_formulism}

Consider a quantum system interacting with the environment, the
Hamiltonian of the whole setup can be generally written as
\begin{eqnarray}
H=H_{0}+\delta H_{I},
\end{eqnarray}%
where $H_{0}$ describes the initial Hamiltonian of the system and
the environment such that $H_{0}|m\rangle =E_{m}|m\rangle $. Here
$\{|m\rangle\}$ is a set of orthogonal basis constructed from the
tensor product of the eigenstate of the system and that of the
environment and $E_{m}$ is the corresponding eigenenergy. $H_{I}$ is
the coupling between the system and the environment with a small
parameter $\delta $ denoting the coupling strength.

Suppose the setup is initially in a state
$|\psi(0)\rangle=|m\rangle$. It is allowed to evolve freely for a
time $\tau$. According to the Schr\"{o}dinger's
equation, the state at time $\tau$ is given by $|\psi(\tau)\rangle=e^{-iH%
\tau}|m\rangle$ and the probability of finding the setup in the
initial state, i.e. the survival probability, is given by
\begin{eqnarray}
P(\tau)&\equiv&\left|p(\tau)\right|^2=\left|\left\langle\psi(0)|\psi(\tau)\right\rangle\right|^2\nonumber\\
&=&\left|\langle m|e^{-i(H_0+\delta H_I)\tau}|m\rangle\right|^2.
\label{eq:survial_prob}
\end{eqnarray}
For small $\delta$, one can expand $p(\tau)$ around $\delta=0$ and
keep terms up to the second order. We have
\begin{eqnarray}
p(\tau)\approx p(\tau)|_{\delta=0}+\delta\frac{\partial p(\tau)}{%
\partial\delta}\Bigg|_{\delta=0}+\frac{\delta^2}{2}\frac{\partial^2p(\tau)}{%
\partial\delta^2}\bigg|_{\delta=0},
\end{eqnarray}
with
\begin{eqnarray}
\frac{\partial
p(\tau)}{\partial\delta}\bigg|_{\delta=0}=(-i\tau)e^{-i\tau
E_m}H_I^{mm},
\end{eqnarray}
and
\begin{eqnarray}
&&\frac{\partial^2p(\tau)}{\partial\delta^2}\Bigg|_{\delta=0}=\nonumber -2e^{-i\tau
E_m}\\ &&\times\sum_{n\neq
m}\left[|H_I^{mn}|^2\frac{1-e^{-i\tau(E_n-E_m)}-i\tau(E_n-E_m)}{(E_n-E_m)^2}\right]\nonumber\\
&&-e^{-i\tau E_m}|H_I^{mm}|^2\tau^2,
\end{eqnarray}
where $H_I^{nm}=\langle n|H_I|m\rangle$. Keeping terms to the lowest
order of $\delta$, Eq. (\ref{eq:survial_prob}) becomes
\begin{eqnarray}
P(\tau)\approx 1-2\delta^2\sum_{n\neq m}\frac{1-\cos[(E_n-E_m)\tau]}{%
(E_n-E_m)^2}|H_I^{mn}|^2,  \label{eq:prob_LE}
\end{eqnarray}
which is just the perturbative form of the Loschmidt echo
\cite{Zhang}. In the short time limit, Eq. (\ref{eq:prob_LE}) gives
\begin{eqnarray}
P(\tau)\approx 1-\delta^2\tau^2\sum_{n\neq m}|H_I^{mn}|^2.
\end{eqnarray}

Now, suppose successive measurements are performed at every time interval $%
\tau $, and the system is allowed to evolve freely in-between
consecutive measurements. The probability of finding the system in
the initial state after a finite duration $\Delta t=Q\tau $, where
$Q$ is the number of measurements, is
\begin{eqnarray}
P(\Delta t)=|P(\tau )|^{Q}  \label{eq:projection}
\approx 1-\Delta
t^{2}\delta ^{2}\frac{\chi }{Q}, \label{eq:del_prob}
\end{eqnarray}%
where
\begin{eqnarray}
\chi \equiv \sum_{n\neq m}|H_{I}^{mn}|^{2}=\langle
m|H_I^2|m\rangle-\langle m|H_I|m\rangle^2, \label{eq:chi}
\end{eqnarray}%
which is the fluctuation in the interaction Hamiltonian. Note that
the projection postulate has been implanted in the first equality in
Eq. (\ref {eq:projection}). From Eq. (\ref{eq:del_prob}), one
realized that there exists a competition between $\chi $ and $Q$.
For a few-level system, since $\chi $ is finite, we have $P(\Delta
t)\rightarrow 1$ under the case of continuous measurements
($Q\rightarrow \infty $). In other words, the system would remain in
the initial state if the measurements are performed continuously.
However, for a quantum many-body system, $\chi$ may not be finite
but goes with some power of the system size. In the thermodynamic
limit, $\chi$ is also infinite and the second term in Eq.
(\ref{eq:del_prob}) could not be simply ignored even if
$Q\rightarrow \infty $. The key motivation of our work is to obtain
the scaling behavior of $\chi$ in quantum many-body systems. From
there, one can then predict the behavior of $\chi$ in the
thermodynamic limit and draw criterions on how large $Q$ should be
in comparison to the system size in order to observe the QZE in the
system.

In the following, we take the one-dimensional transverse-field Ising
model and the LMG model coupled to a central qubit as examples.
Using Eq. (\ref{eq:chi}), we calculated $\chi$ explicitly and
extracted its size dependence. One can then determine the criteria
on how large $Q$ should be in order to observe the QZE in the
models.

\section{One-dimensional transverse-field Ising model}
\label{sec_Ising}

The Hamiltonian of the one-dimensional transverse-field Ising model
reads
\begin{eqnarray}
H_{\mathrm{Ising}}=-\sum_{i}(\sigma _{i}^{z}\sigma
_{i+1}^{z}+\lambda \sigma _{i}^{x}), \label{eq:Ising_H}
\end{eqnarray}
where $\sigma _{i}^{q}$, with $q=x,y,z$, is the Pauli matrix. It
describes a chain of spins interacting with the nearest neighbors in
the $z$ direction, while all the spins are subjected to a transverse
external field with strength $\lambda $ in the $x$ direction.

The Hamiltonian in Eq. (\ref{eq:Ising_H}) can be diagonalized
exactly following three standard transformations \cite{trans}: i.e.
\emph{Jordan-Wigner transformation}, \emph{Fourier transformation},
and \emph{Bogoliubov transformation}. Then the Hamiltonian in Eq.
(\ref{eq:Ising_H}) can be written into an diagonal form of
$H_{\mathrm{Ising}}=\sum_{k}\varepsilon _{k}(d_{k}^{\dagger
}d_{k}-1/2),$ where $\varepsilon _{k}=2\sqrt{1+\lambda ^{2}-2\lambda
\cos k}$ is the single quasi-particle energy. $d_k$ and
$d_k^{\dagger}$ is the fermionic annihilation and creation operator
respectively.

Now, consider a central qubit interacting transversely with the
Ising model. The corresponding Hamiltonian reads
\begin{eqnarray}
H& =&H_{\mathrm{Ising}}-\delta \sum_{i}|e\rangle \langle e|\sigma _{i}^{x},\nonumber\\
& =&H_{\mathrm{Ising}}|g\rangle \langle g|+(H_{\mathrm{Ising}}+\delta
H_{I})|e\rangle \langle e|,
\end{eqnarray}%
where $H_{\mathrm{Ising}}$ is given by Eq. (\ref{eq:Ising_H}), and $%
H_{I}=-\sum_{i}\sigma _{i}^{x}$. $|g\rangle $ and $|e\rangle $
denotes the ground state and the excited state of the central qubit,
respectively. The Loschmidt echo in this system has been
investigated by Quan and his collabrators \cite{Quan}. They found
that the decay of Loschmidt echo is greatly enhanced at the critical
point of the Ising model. In the following, we will follow similar
approach as used by Quan et al., but focusing on the short-time
limit of the Loschmidt echo.

Without loss of generality, assume the Ising model is initially in
some state $|\phi (0)\rangle $ and the central qubit is in a superposition state $%
c_{g}|g\rangle +c_{e}|e\rangle $, where the coefficients satisfie $%
|c_{g}|^{2}+|c_{e}|^{2}=1$. The initial state of the whole setup is thus $%
|\psi (0)\rangle =|\phi (0)\rangle \otimes (c_{g}|g\rangle
+c_{e}|e\rangle )$. The evolution of the Ising chain splits into two
branches and the state vector of the setup at a later time $\tau $
is given by $|\psi (\tau )\rangle =c_{g}e^{-iH_{\mathrm{Ising}}\tau
}|\phi (0)\rangle \otimes |g\rangle
+c_{e}e^{-i(H_{\mathrm{Ising}}+\delta H_{I})\tau }|\phi (0)\rangle
\otimes |e\rangle $.

\begin{figure}[tbp]
\centering
\includegraphics[width=8cm]{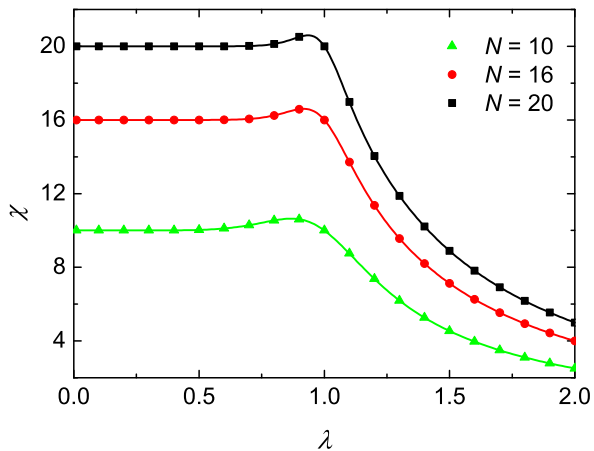}
\caption{$\protect\chi$ as a function of $\protect\lambda$ for the
one-dimensional transverse-field Ising model coupled to a central
qubit for system size $N=10$, $16$ and $20$. The symbols denote data
obtained from the numerical exact diagonalization while the lines
are obtained from the analytical expression of $\chi$ in Eq.
(\ref{eq:chi_Ising}). The analytical results agree with the
numerical data. } \label{fig:Ising_chi_check}
\end{figure}

For simplicity, suppose the Ising chain is initially in the ground
state $|G\rangle $, and the central qubit is in the excited state, i.e. $%
c_{g}=0$ and $c_{e}=1$. The survival probability of the system as
given in Eq. (\ref{eq:survial_prob}) becomes
\begin{equation}
P(\tau )=|\langle G|e^{-i(H_{\mathrm{Ising}}+\delta
H_{I})\tau}|G\rangle |^{2}.
\end{equation}
Note that in the eigen-basis of the diagonalized Hamiltonian $|n_k\rangle$ such that $%
d_k^{\dagger}d_k|n_k\rangle=n_k|n_k\rangle$, where $n_k=0$ or 1, the
ground state of the Ising model can be written as
$|G\rangle=\prod_{k>0}|0_k, 0_{-k}\rangle$.  

By performing the Jordan-Wigner transformation, Fourier
transformation, and the Bogoliubov transformation, $H_I$ can also be
transformed into the eigen-basis $\{|n_k\rangle\}$ as
\begin{eqnarray}
&&H_I=-N \\\nonumber &&+\sum_k\left[2(\cos\theta_k)d_k^{\dagger}d_k+i(\sin\theta_k)(d_k^{\dagger}d_{-k}^{\dagger}-d_{-k}d_k)\right],
\label{eq:Ising_H_I}
\end{eqnarray}
where $\theta_k$ is the Bogoliubov angle satisfying
$\tan\theta_k=\sin k(\cos k-\lambda)^{-1}$. Using Eq.
(\ref{eq:chi}), we obtain
\begin{eqnarray}
\chi=\sum_{k>0}\frac{4\sin^2k}{1+\lambda^2-2\lambda\cos k}.
\label{eq:chi_Ising}
\end{eqnarray}

\begin{figure}[tbp]
\centering
\includegraphics[width=8cm]{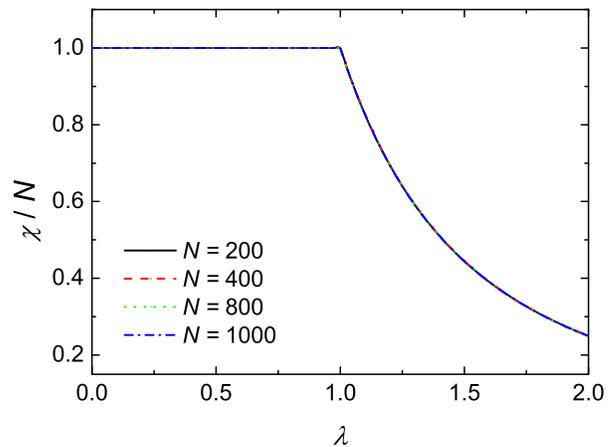}
\caption{The normalized $\protect\chi$ as a function of $\lambda$
for the one-dimensional transverse-field Ising model coupled to a
central qubit. The curve for different system sizes collapses to a
single one.} \label{fig:Ising_avg_chi}
\end{figure}

Next, we would like to investigate the size dependence of $\chi$.
Assume $N$ is even, we introduce the density of states in $k$-space
in the limit of $N\rightarrow \infty $. For $\lambda =1$, one gets
\begin{eqnarray}
\chi=\frac{N}{\pi }\int_{\pi /N}^{(N-1)\pi /N}\frac{\sin
^{2}k}{1-\cos k}dk \sim N.  \label{eq:lamda1}
\end{eqnarray}%
For $\lambda \neq 1$, let $0\leq x\leq 1/2$ and $0\leq k=2\pi x\leq \pi $, then Eq. (\ref%
{eq:chi_Ising}) becomes
\begin{equation}
\chi =2N\int_{0}^{1}\frac{\sin^{2}2\pi x}{1+\lambda ^{2}-2\lambda
\cos2\pi x}dx.
\end{equation}%
Let $z=e^{i2\pi x}$, so $\sin (2\pi x)=\frac{1}{2i}(z-\frac{1}{z})$ and $%
\cos (2\pi x)=\frac{1}{2}(z+\frac{1}{z})$, the integral becomes the
contour integral along a unit circle on the complex plane:
\begin{equation}
\chi =\frac{N}{4\pi i}\oint \frac{(z^{2}-1)^{2}}{%
z^{2}(z-\lambda )(\lambda z-1)}dz,
\end{equation}
which can be easily evaluated using the residue theorem. One then
obtains
\begin{equation}
\chi =\left\{
\begin{array}{ll}
N/\lambda ^{2}\sim N &\quad\textrm{for $\lambda >1$}
\\
N\sim N & \quad \textrm{for $\lambda <1$}.
\end{array}
\right.   \label{eq:lam_ne_ising}
\end{equation}


To check the validity of the result obtained above, one may consider
the limiting case where $\lambda\rightarrow 0$. In this limit, the
ground state of the system is a superposition of two polarized
wavefunctions, i.e. $\frac{1}{\sqrt
2}(|\uparrow\uparrow\cdots\uparrow\rangle+|\downarrow\downarrow\cdots\downarrow\rangle)$.
Noting that the interaction Hamiltonain $H_I$ only flips a
particular spin at site $i$. This results in a wavefunction
orthogonal to the original one. The second term in Eq.
(\ref{eq:chi}) is thus zero. For the first term, one can also easily
verify that gives $N$ using the property of the Pauli's matrices.
Therefore, one obtains $\chi=N$ as $\lambda\rightarrow 0$. On the
other hand, form the analytical expression in Eq.
(\ref{eq:chi_Ising}), we also obtain the same result for $\lambda=0$
as shown in Eq. (\ref{eq:lam_ne_ising}).

Furthermore, we also perform the numerical exact diagonalization and
calculated $\chi$ using Eq. (\ref{eq:chi}). Fig.
\ref{fig:Ising_chi_check} shows a plot of $\chi$ as a function
$\lambda$ for $N=10$, $16$ and $20$. The lines obtain from the
analytical expression of $\chi$ agree with the numerical data
(represented by symbols).

Fig. \ref{fig:Ising_avg_chi} shows a plot of $\chi$ given by the
analytical expression in Eq. (\ref {eq:chi_Ising}) normalized over
the system size as a function of $\lambda$. The curve for different
system sizes collapse to a single one, i.e. $\chi\sim N$. This
echoes the result obtained from the scaling analysis above.

Returning to Eq. (\ref{eq:del_prob}), we see that in order to
observe QZE in the Ising model, $Q\gg \xi N$ where $\xi $ is a
constant independent of the system size. In the thermodynamic limit,
in which $N$ is of order $10^{23}$, $Q$ has to be at least of this
order or else the QZE breaks down.

\begin{figure}[tbp]
\centering
\includegraphics[width=8cm]{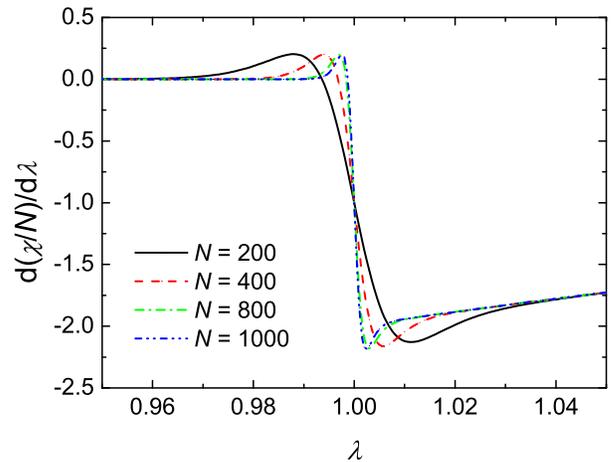}
\caption{The first derivative of the normalized $\protect\chi$ with
respect to $\lambda$ as a function of $\lambda$ near the vicinity of
the critical point for the one-dimensional transverse-field Ising
model coupled to a central qubit. $d(\chi/N)/d\lambda$ changes from
positive to negative across the critical point of the Ising model.
The change is faster for a larger system size. The curve for
different system sizes coincide exactly at $\lambda=1$.}
\label{fig:Ising_davg_chi}
\end{figure}

\begin{figure}[tbp]
\centering
\includegraphics[width=8cm]{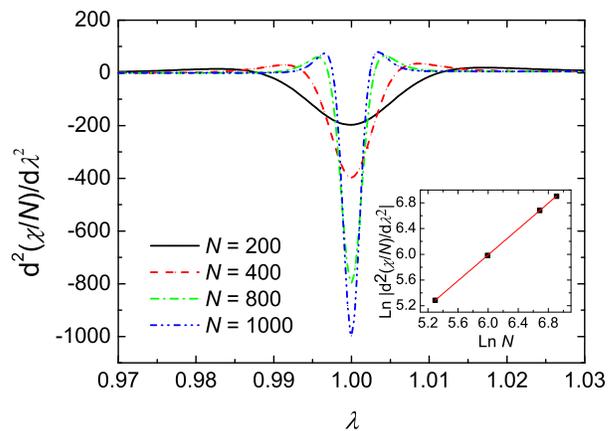}
\caption{The second derivative of the normalized $\protect\chi$ with
respect to $\lambda$ as a function of $\lambda$ near the vicinity of
the critical point for the one-dimensional transverse-field Ising
model coupled to a central qubit. The inset shows the scaling
behavior of the minimum of $d^2(\chi/N)/d\lambda^2$. The slope of
the straight line is $1.007\pm0.001$.} \label{fig:Ising_ddavg_chi}
\end{figure}

From Fig. \ref{fig:Ising_avg_chi}, one can also realize that the
normalized $\chi$ changes manifestly across the point at
$\lambda=1$, which is the critical point where a quantum phase
transition(QPT) of the Ising model takes place \cite{trans}.
Differentiating Eq. (\ref{eq:chi_Ising}) with respect to $\lambda$,
one obtains
\begin{eqnarray}
\frac{d\chi}{d\lambda}=\sum_{k>0}\frac{8(\cos
k-\lambda)\sin^2k}{(1+\lambda^2-2\lambda\cos k)^2}.
\label{eq:ising_dchi}
\end{eqnarray}
A plot of Eq. (\ref{eq:ising_dchi}) normalized over the system size
as a function of $\lambda$ in the vicinity of the critical point is
shown in Fig. \ref{fig:Ising_davg_chi}. $d(\chi/N)/d\lambda$ changes
from positive to negative as the system goes across the critical
point from below. This change becomes sharper and sharper as the
system size increases. Fig. \ref{fig:Ising_ddavg_chi} shows the
second derivative of the normalized $\chi$ as a function of
$\lambda$ in the vicinity of the critical point. One can see that
$d^2(\chi/N)/d\lambda^2$ exhibits a minimum at the critical point.
The magnitude of this minimum is in fact scales with $N$, as shown
in the inset of Fig. \ref{fig:Ising_ddavg_chi}.

The size dependence of $d^2(\chi/N)/d\lambda^2$ can also be obtained
analytically. Differentiate Eq. (\ref{eq:ising_dchi}) with respect
to $\lambda$, one obtains
\begin{eqnarray}\label{eq:ising_ddchi}
&&\frac{d^2\chi}{d\lambda^2}=  \\ \nonumber &&\sum_{k>0}\left[\frac{32(\lambda-\cos
k)^2\sin^2 k}{(1+\lambda^2-2\lambda\cos k)^3}
-\frac{8\sin^2k}{(1+\lambda^2-2\lambda\cos
k)^2}\right].
\end{eqnarray}
For $\lambda=1$,
\begin{eqnarray}
\frac{d^2\chi}{d\lambda^2}&=&\sum_{k>0}\left[\frac{4\sin^2k}{1-\cos
k}-\frac{2\sin^2k}{(1-\cos k)^2}\right],\nonumber \\
&=&\frac{N}{\pi}\int^{(N-1)\pi/N}_{\pi/N}\left[2(1+\cos
k)-\frac{1+\cos k}{1-\cos k}\right]dk, \nonumber \\ &\sim &N^2.
\label{eq:ising_ddchi_lam1}
\end{eqnarray}
We then have $d^2(\chi/N)/d\lambda^2\sim N$ which diverges in the
thermodynamic limit.

Furthermore, from Fig. \ref{fig:Ising_davg_chi}, we see that there
exists a fixed point at $\lambda=1$ where $d(\chi/N)/d\lambda$ for
various system size coincide. This can also be shown analytically
from Eq. (\ref{eq:ising_dchi}). For $\lambda=1$, Eq.
(\ref{eq:ising_dchi}) gives
\begin{eqnarray}
\frac{d(\chi)}{d\lambda}&=&\sum_{k>0}\frac{2\sin^2k}{1-\cos k}\nonumber \\
&=&\frac{N}{\pi}\int^{(N-1)\pi/N}_{\pi/N}(1+\cos k)dk \nonumber \\&=&-N,
\end{eqnarray}
in which the average over system size gives $-1$. This fixed point
reflects some kind of symmetry is hidden in the model and further
exploration would be of interest.

\section{Lipkin-Meshkov-Glick model}
\label{sec_LMG}

\begin{figure}[bpt]
\centering
\includegraphics[width=8cm]{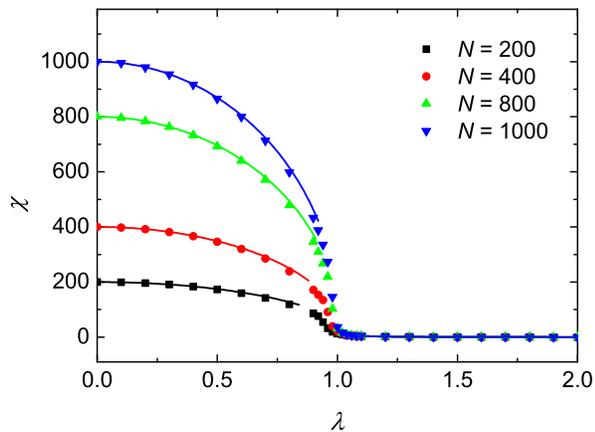}
\caption{$\protect\chi$ as a function of $\protect\lambda$ for the
LMG model coupled to a central qubit with system sizes $N=800$ and
1000 ($\gamma=0$). The dots denote data obtained from the numerical
exact diagonalization while the smooth lines are obtained from the
analytical expression of $\chi$ in Eq. (\ref{LMG_chi_N0}) and
(\ref{LMG_chi_N}). The analytical results agree with the numerical
data except for points close to $\lambda=1$.}
\label{fig:LMG_chi_check}
\end{figure}

 The Hamiltonian of the LMG model reads
\begin{equation}
H_{\mathrm{LMG}}=-\frac{1}{N}\sum_{i<j}\left(\sigma _{i}^{x}\sigma
_{j}^{x}+\gamma \sigma _{i}^{y}\sigma _{j}^{y}\right)-\lambda
\sum_{i}\sigma _{i}^{z}.  \label{eq:LMG_sigma}
\end{equation}
The Hamiltonian describes a cluster of spins mutually interacting
with each other in the $xy$ plane. $\gamma $ is the parameter
indicating the anisotropy in the interaction of the model. All the
spins are embedded in an external field along the $z$ direction. The
prefactor $1/N$ is to ensure a finite energy per spin in the
thermodynamic limit. Introducing $S_{\kappa }=\sum_{i}\sigma
_{i}^{\kappa}/2$, where $\kappa =x,y,z$, and $S_{x}=(S_{+}+S_{-})/2$ and $%
S_{y}=(S_{+}-S_{-})/2i$, Eq. (\ref{eq:LMG_sigma}) can be rewritten
as
\begin{eqnarray}
H_{\mathrm{LMG}}&=& -2\lambda S_{z}-\frac{1}{2N}(1-\gamma
)\left(S_{+}^{2}+S_{-}^{2}\right) \nonumber \\ &&  -\frac{1}{N}(1+\gamma
)\left(S^{2}-S_{z}^{2}-\frac{N}{2}\right). \label{eq:LMG_S}
\end{eqnarray}
In the large $N$ limit and considering the low excitation spectrum,
the above Hamiltonian can be diagonalized through a semi-classical
approach as in the following\cite{Vidal}:

\begin{figure}[bpt]
\centering
\includegraphics[width=8cm]{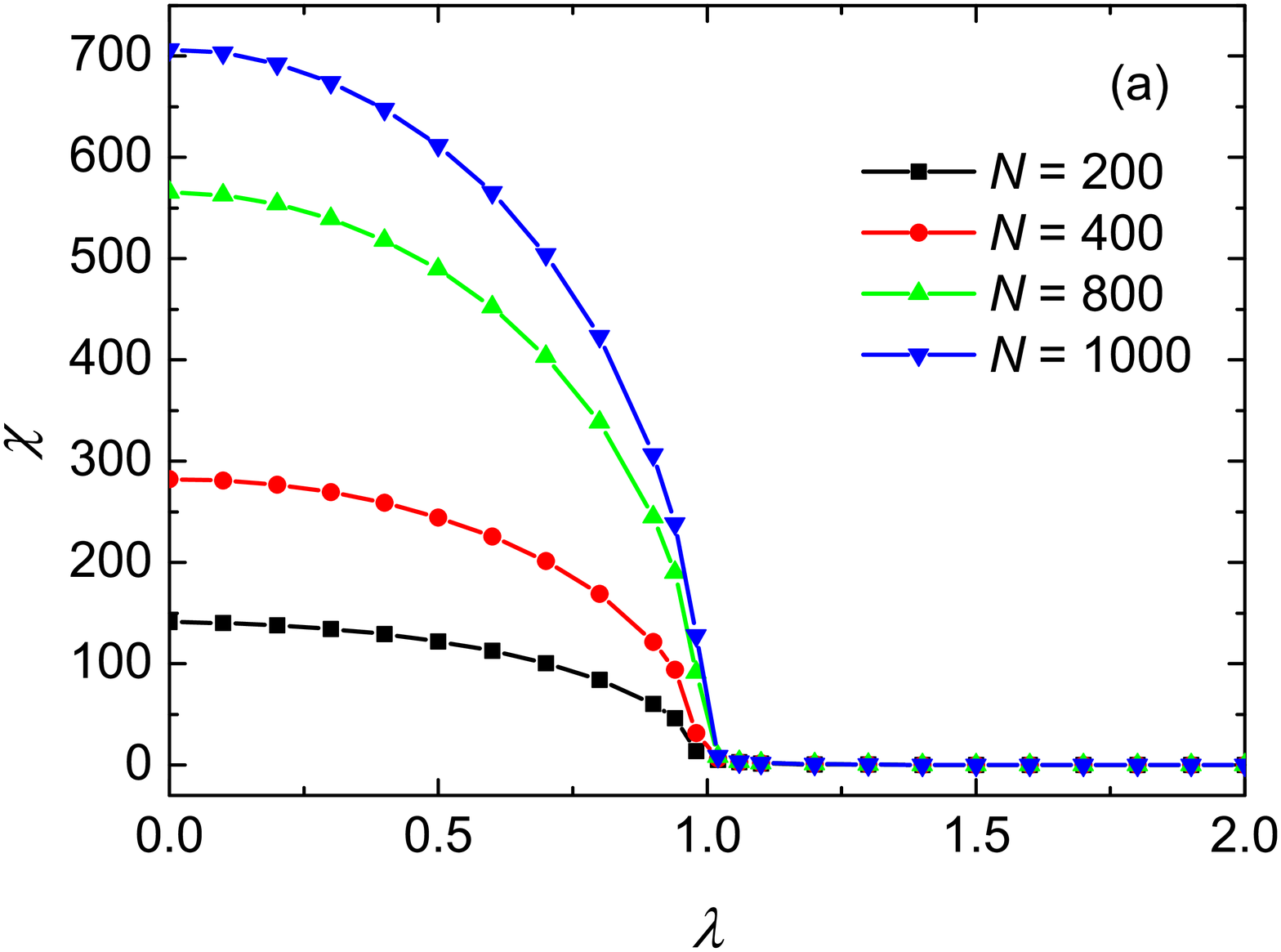}\\
\includegraphics[width=8cm]{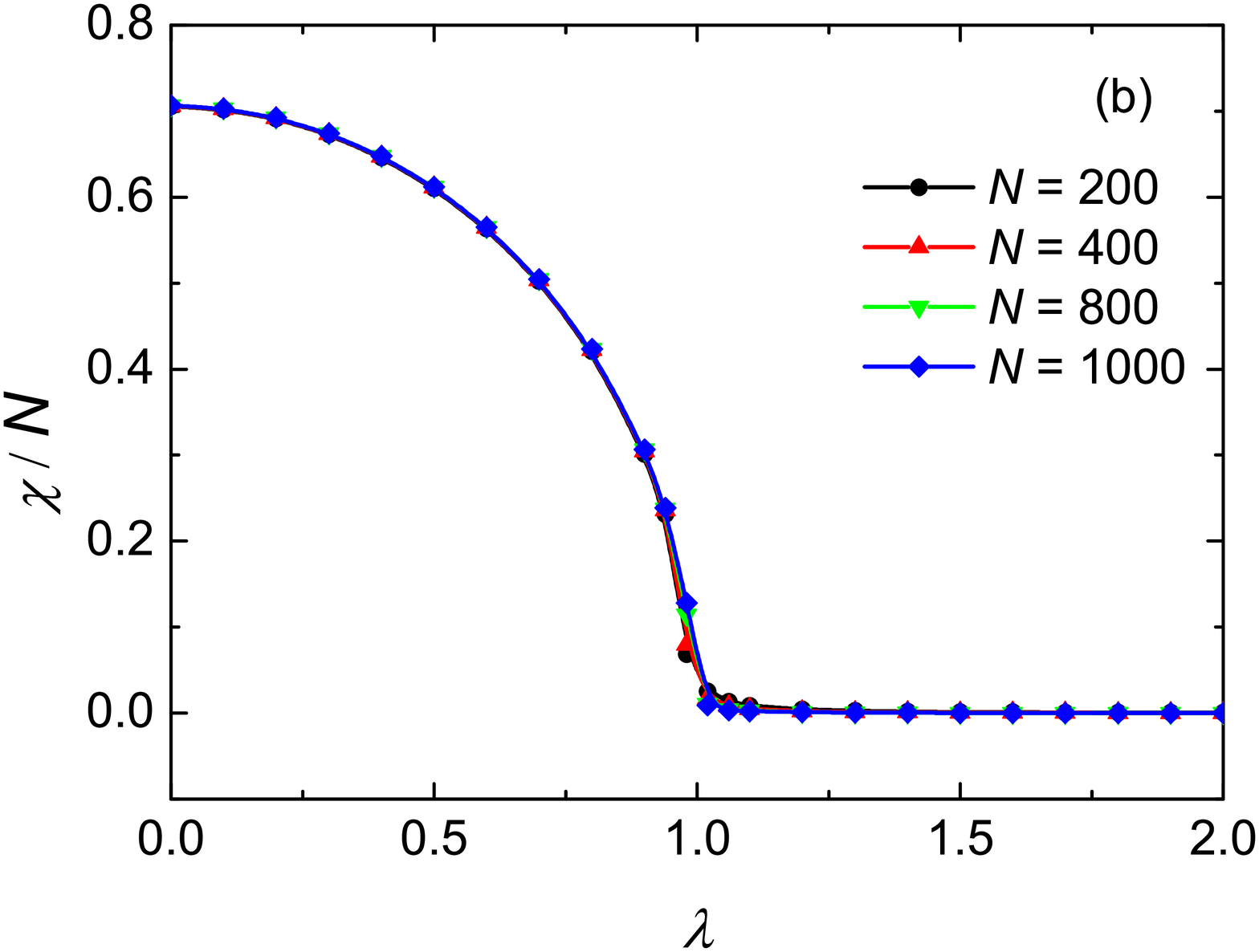}
\caption{(a) $\protect\chi$ as a function of $\lambda$ (b) The
normalized $\protect\chi$ as a function of $\protect\lambda$ for
different system sizes for the LMG model coupled to a central qubit
($\protect\gamma=0.5$). The normalized $\chi$ for various system
sizes collapse to a single curve.} \label{fig:LMG_chi_ED}
\end{figure}

(i) Perform a rotation of the spin operators around the $y$ axis in
order to bring the $z$ axis along the semiclassical magnetization.
This is done by
\begin{eqnarray}
\left( {\begin{array}{*{20}c}
   {S_x }  \\
   {S_y }  \\
   {S_z }  \\
\end{array}} \right) = \left( {\begin{array}{*{20}c}
   {\cos \alpha } & 0 & {\sin \alpha }  \\
   0 & 1 & 0  \\
   { - \sin \alpha } & 0 & {\cos \alpha }  \\
\end{array}} \right)\left( {\begin{array}{*{20}c}
   {\tilde S_x }  \\
   {\tilde S_y }  \\
   {\tilde S_z }  \\
\end{array}} \right)
,  \label{rotate}
\end{eqnarray}%
where $\alpha =0$ for $\lambda >1$ so that
$\mathbf{S}=\mathbf{\widetilde{S}}$, and $\alpha =\cos ^{-1}\lambda
$ for $\lambda <1$.

(ii) \emph{Holstein-Primakoff transformation} which transforms the
spin operators into bosonic operators by
\begin{eqnarray}
S_{z}& =S-a^{\dagger }a=N/2-a^{\dagger }a, \\
S_{+}& =(2S-a^{\dagger}a)^{1/2}a=\sqrt{N}(1-a^{\dagger
}a/N)^{1/2}a,\\
S_{-}& =a^{\dagger
}(2S-a^{\dagger}a)^{1/2}=\sqrt{N}a^{\dagger}(1-a^{\dagger
}a/N)^{1/2},  \label{HP}
\end{eqnarray}%
where $a$ and $a^{\dagger }$ are bosonic annihilation and creation
operators satisfying the commutation relation $[a,a^{\dagger }]=1$.
The second equality in above equations hold as a result of
$[H,S^2]=0$ and only the low energy spectrum is considered.

(iii)\emph{Bogoliubov transformation}: After the above
transformations, we arrive at a quadratic Hamiltonian and one can
easily diagonalize it using the Bogoliubov transformation:
\begin{eqnarray}
a =\sinh \left(\frac{\theta}{2}\right)b^{\dagger}+\cosh
\left(\frac{\theta}{2}\right)b,
\\a^{\dagger}=\cosh \left(\frac{\theta}{2}\right)b^{\dagger }+\sinh \left(\frac{\theta}{2}\right)b,
\label{BT_LMG}
\end{eqnarray}
where $b$ and $b^{\dagger }$ are also bosonic operators satisfying
the same commutation relation as that of $a$ and $a^{\dagger }$.
Following the above transformations, one could finally diagonalize
the Hamiltonian in Eq. (\ref{eq:LMG_S}) into
$H_{\mathrm{LMG}}=\epsilon_{0}N+\epsilon_1-\Delta/2+\sqrt{\Delta^2-4\Gamma^2}(b^{\dagger}b+1/2)$,
where for $\lambda<1$,
\begin{eqnarray}
\left\{\begin{array}{lcl}
\epsilon_0&=&-\displaystyle\frac{1+\lambda^2}{2}\\
\epsilon_1&=&\displaystyle\frac{1-\lambda^2}{2} \\
\Delta&=&2-\lambda^2-\gamma\\
\Gamma&=&\displaystyle\frac{\gamma-\lambda^2}{2} \end{array}\right..
\label{parameters1}
\end{eqnarray}
For $\lambda>1$,
\begin{eqnarray}
\left\{\begin{array}{lcl}
\epsilon_0&=&-\lambda\\
\epsilon_1&=&0\\
\Delta&=&2\lambda-1-\gamma\\
\Gamma&=&-\displaystyle\frac{1-\gamma}{2}
\end{array}\right..\end{eqnarray}\label{parameter2}
The diagonalization condition is given by
\begin{eqnarray}
\tanh\theta=-\frac{2\Gamma}{\Delta},  \label{LMG_tan}
\end{eqnarray}
from which we also have
\begin{eqnarray}
\cosh\theta=\frac{\Delta}{\sqrt{\Delta^2-4\Gamma^2}},
\quad\sinh\theta=\frac{-2\Gamma}{\sqrt{\Delta^2-4\Gamma^2}}.
\label{LMG_sin}
\end{eqnarray}

Similar to the case of the Ising model in the previous section,
let's consider a central qubit interacting transversely with the LMG
model. The corresponding Hamiltonian reads
\begin{eqnarray}
H=H_{\mathrm{LMG}}-\delta \sum_{i}|e\rangle \langle
e|\sigma_{i}^{z},
\end{eqnarray}
where $H_{\mathrm{LMG}}$ is given by Eq. (\ref{eq:LMG_sigma}).
Again, suppose the LMG model is initially in the ground state
$|G\rangle$, and the central qubit is in the excited state.
Following the same argument as in the Ising model, the survival
probability of the system remain in the initial state in Eq.
(\ref{eq:survial_prob}) becomes
\begin{eqnarray}
P(\tau )=|\langle G|e^{-i(H_{\mathrm{LMG}}+\delta
H_{I})\tau}|G\rangle |^{2},
\end{eqnarray}
where $H_{I}=-\sum_{i}\sigma _{i}^{z}$. Similarly, in the
eigen-basis $\{|n\rangle\}$ such that
$b^{\dagger}b|n\rangle=n|n\rangle$, where now $n=0, 1, 2\cdots$,
note that ground state of the LMG model is just
$|G\rangle=|0\rangle$. Following the diagonalization procedure
mentioned above, $H_I$ can also be expressed in the eigen-basis
$\{|n\rangle\}$. For $\lambda>1$,
\begin{eqnarray}
H_I&=&N+1-\cosh\theta-2\cosh(\theta)b^{\dagger}b \nonumber \\ && -\sinh(\theta)(b^{\dagger
2}+b^2). \label{eq:LMG_H_Ilarger1}
\end{eqnarray}
Using Eq. (\ref{eq:chi}), (\ref{LMG_sin}), and
(\ref{eq:LMG_H_Ilarger1}), one obtains
\begin{eqnarray}
\chi=2\sinh^2\theta\sim N^0.  \label{LMG_chi_N0}
\end{eqnarray}
For $\lambda<1$,
\begin{eqnarray}
H_I&=&-\lambda(N-\cosh(\theta)+1)\nonumber \\ && +2\cosh(\theta)b^{\dagger}b+\lambda\sinh(\theta)\left(b^{\dagger2}+b^2\right)\\ \nonumber
&&+\sqrt{N(1-\lambda^2)}(b^{\dagger}+b)(\cosh(\theta/2)+\sinh(\theta/2)).
\label{eq:LMG_H_Ismaller1}
\end{eqnarray}
Using Eq. (\ref{eq:chi}), (\ref{LMG_sin}), and
(\ref{eq:LMG_H_Ismaller1}), one has
\begin{eqnarray}
\chi&=&2\lambda\sinh^2\theta \nonumber \\ && +\lambda\sqrt{N(1-\lambda^2)}\sinh\theta(\sqrt{\cosh\theta+1}+\sqrt{\cosh\theta-1})\nonumber \\
&&+(1-\lambda^2)N(\cosh\theta+\sinh\theta)\nonumber \\
&\sim &N.  \label{LMG_chi_N}
\end{eqnarray}

\begin{figure}[bpt]
\centering
\includegraphics[width=8cm]{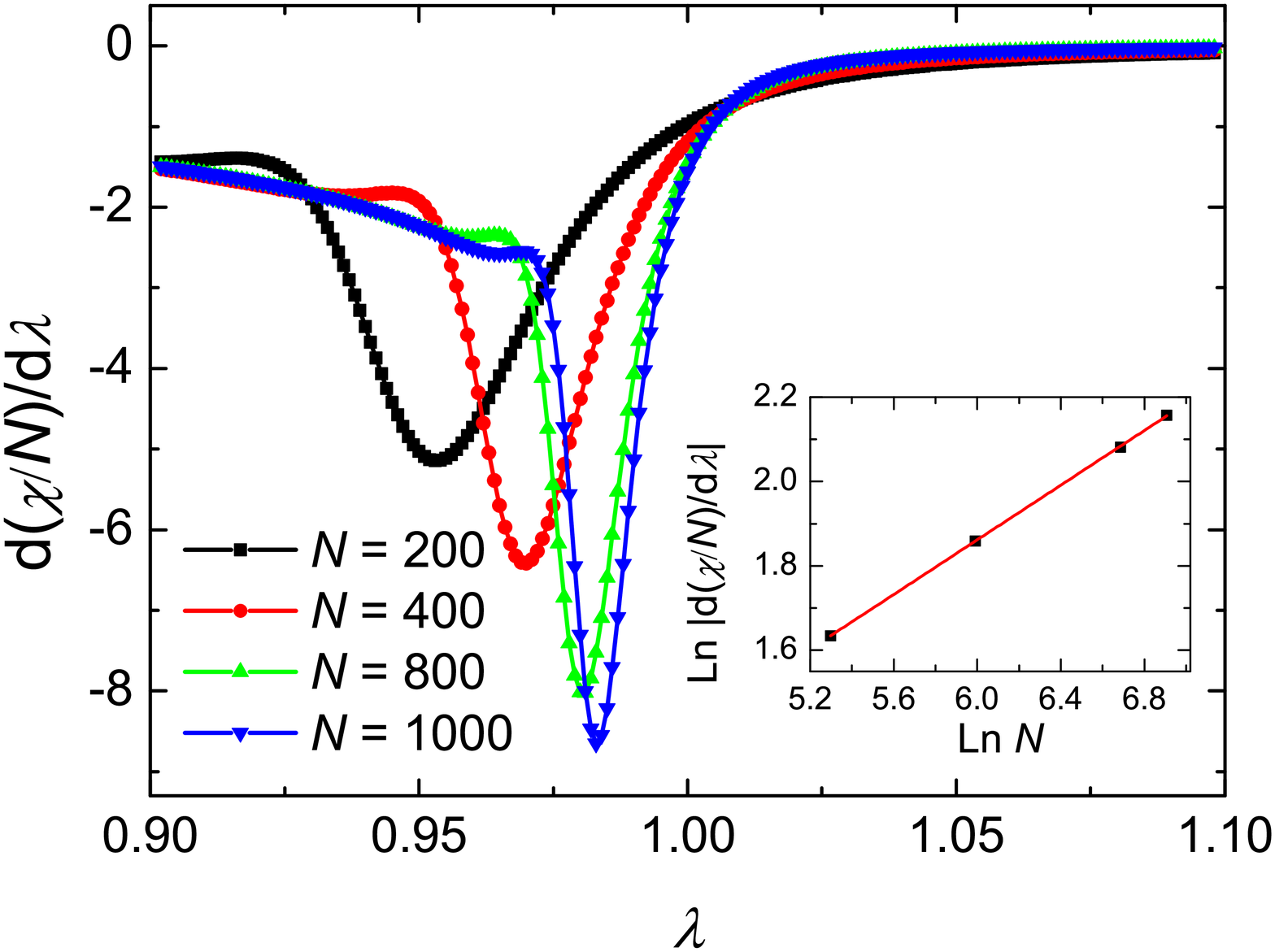}
\caption{First derivative of the normalized $\protect\chi$ with
respect to $\lambda$ as a function of $\protect\lambda$ in the
vicinity of the critical point for the LMG model coupled to a
central qubit ($\protect\gamma=0.5$). The inset shows the scaling
behavior of the minimum of $d(\chi/N)/d\lambda$. The slope of the
straight line gives $0.324\pm0.001.$} \label{fig:LMG_dchi_gamma05}
\end{figure}

\begin{figure}[bpt]
\centering
\includegraphics[width=8cm]{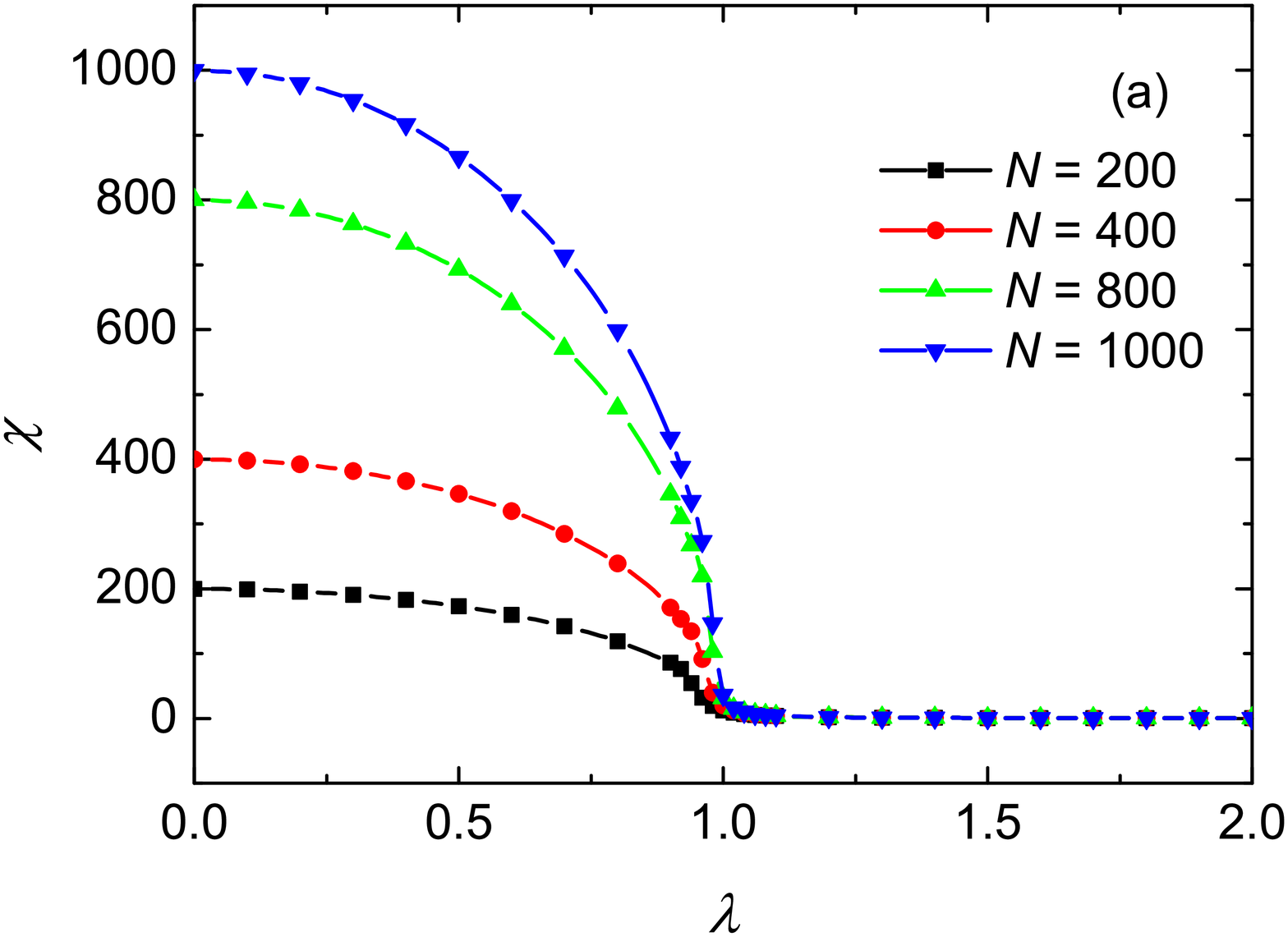}\\
\includegraphics[width=8cm]{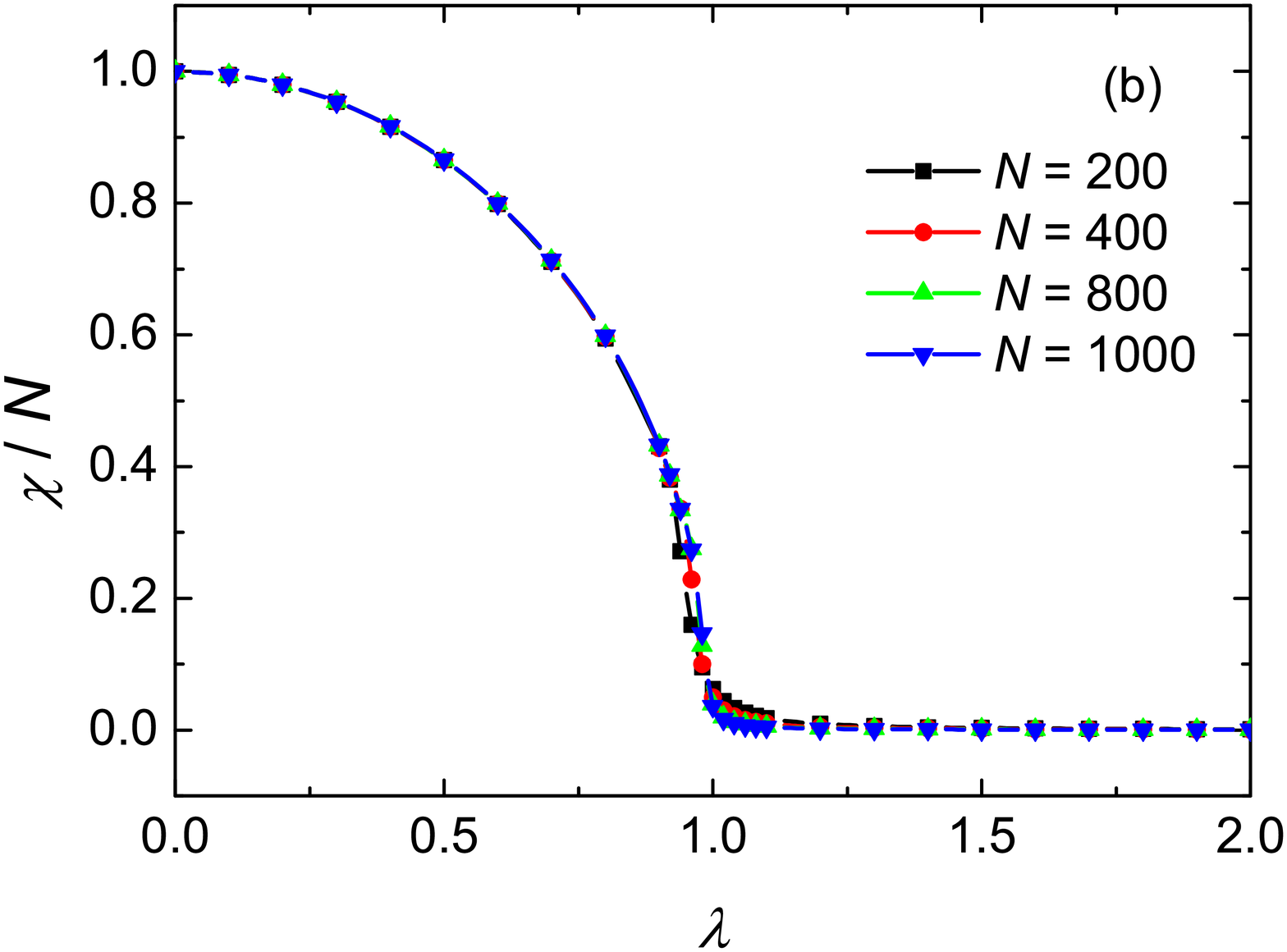}
\caption{(a) $\protect\chi$ (b) The normalized $\protect\chi$ for
different system sizes as a function of $\protect\lambda$ for the
LMG model coupled to a central qubit ($\protect\gamma=0$). The
normalized $\chi$ for various system sizes collapse to a single
curve.} \label{fig:LMG_chi_gamma00}
\end{figure}

\begin{figure}[bpt]
\centering
\includegraphics[width=8cm]{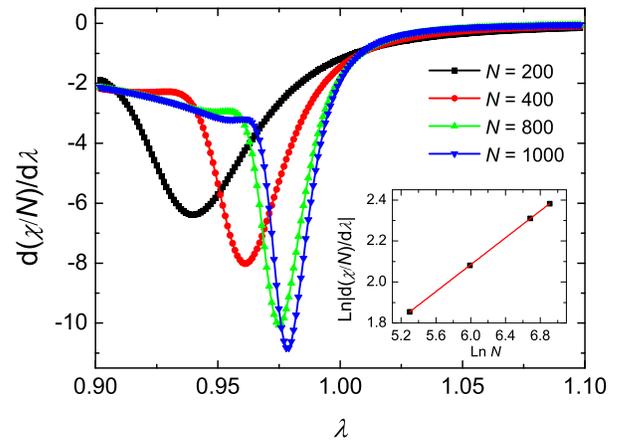}
\caption{First derivative of the normalized $\protect\chi$ with
respect to $\lambda$ as a function of $\protect\lambda$ in the
vicinity of the critical point for the LMG model coupled to a
central qubit ($\protect\gamma=0$). The inset shows the scaling
behavior of the minimum of $d(\chi/N)/d\lambda$. The slope of the
straight line gives $0.328\pm0.001.$} \label{fig:LMG_dchi_gamma00}
\end{figure}

Again, to check the validity of the analytical result, let's first
consider the limiting case where $\gamma=0$ and $\lambda\rightarrow
0$. In this limit, the Hamiltonian as given by Eq.
(\ref{eq:LMG_sigma}) is similar to that of the Ising model except
the interaction is over all spins here. Following similar argument
as that for the Ising model, $\chi$ obtained from Eq. (\ref{eq:chi})
also gives $N$. On the other hand, from Eq. (\ref{parameters1}) and
(\ref{LMG_sin}), one can easily see that Eq. (\ref{LMG_chi_N}) also
gives $N$. The two results are consistent.

Moreover, the analytical result is also compared with the data from
the numerical exact diagonalization. As shown in Fig.
\ref{fig:LMG_chi_check}, the two data agree with each other except
for $\lambda$ close to $1$. This invalidity of the analytical result
for finite system sizes near the critical point is due to the fact
that we have only kept terms up to the lowest order in the $1/N$
expansion of the HP transformation. To obtain more accurate results
near the critical point, one have to go to higher order terms in the
$1/N$ expansion of HP transformation \cite{Vidal} or through the
analysis of the Majorana polynomial roots \cite{Vidal2}. In this
report, our focus is on the scaling behavior of $\chi$ within the
same phase of the model. The analytical analysis presented so far
remains valid as long as one stays in the same phase.

Fig. \ref{fig:LMG_chi_ED} shows a plot of $\chi$ obtained from exact
diagonalization as a function of $\lambda$ for the case of
$\gamma=0.5$. $\chi$ shows different scaling behavior on two sides
of $\lambda=1$. This is the quantum critical point of the LMG model.
For $\lambda>1$, the system is in the polarized phase and spins are
polarized along the $z$-direction. In this phase, $\chi$ for
different system sizes collapse to a single curve indicating that
$\chi\sim N^0$. For $\lambda<1$, the model is in the symmetry-broken
phase. The spin-spin interaction in the model in the $xy$ plane
dominates. The normalized $\chi$ for different size collapse into a
single curve and one has $\chi\sim N$. These results are consistent
with the scaling analysis using the analytical expression of $\chi$
as shown in Eq. (\ref{LMG_chi_N0}) and Eq. (\ref{LMG_chi_N}),
respectively.

To interpret the results obtained, we argue that in order to observe
the QZE in the symmetry broken phase, the number of measurements $Q$
has to be of comparable order of $N$. In a thermodynamic system,
this condition is hardly realized. However, in the polarized phase
where $\chi$ is independent of the system size, the QZE is much more
easier to be observed in comparing to that in the symmetry broken
phase. Physically, this may be understood by the fact that in the
case where $\lambda>1$, the ground state of the model is given by
the configuration in which spins are already fully polarized along
the external field.

Moreover, from Fig. \ref{fig:LMG_chi_ED}, we also see that $\chi$
shows some abnormal behavior at the quantum critical point of the
LMG model. Fig. \ref{fig:LMG_dchi_gamma05} shows a plot of the first
derivative of the normalized $\chi$ with respect to $\lambda$ near
the vicinity of the critical point. $d(\chi/N)/d\lambda$ exhibits a
minimum near $\lambda=1$ and this minimum's amplitude increases as
the system size increases. From the inset of the figure, one finds
$d(\chi/N)/d\lambda\sim N^{0.324\pm0.001}$. In the thermodynamic
limit where $N\rightarrow\infty$, $d(\chi/N)/d\lambda$ is expected
to be divergent at the critical point.

Fig. \ref{fig:LMG_chi_gamma00} and \ref{fig:LMG_dchi_gamma00} shows
the numerical exact diagonalization result of $\chi$ and
$d(\chi/N)/d\lambda$ as a function of $\lambda$ respectively for the
case of $\gamma=0$. The scaling behavior of $\chi$ obtained here is
the same as the case for $\gamma=0.5$. From the inset of Fig.
\ref{fig:LMG_dchi_gamma00}, we also have $d(\chi/N)/d\lambda\sim
N^{0.328\pm0.001}$, which is consistent with the case for
$\gamma=0.5$ up to two digits.

\section{summary}
\label{sec_sum}

In this report, we have investigated the QZE from the viewpoint of
condensed matter physics. We have obtained the scaling behavior of
$\chi$ in two analytically solvable systems, namely the
one-dimensional transverse-field Ising model and the LMG model. We
have found that in the Ising model, the frequency of the projective
measurement should be of comparable order to that of the system
sizes in order to observe the QZE, and similar conclusion is
obtained for the case of the symmetry broken phase in the LMG model;
however for the polarized phase of the LMG model, the QZE can be
easily observed via frequent measurements. Since we know that the
size of the system in the thermodynamic limit is of order $10^{23}$,
in reality it is almost impossible to have such a high frequency of
projective measurement by the current technology. In this sense, we
can safely argue that under certain circumstances, the QZE may
breaks down in the thermodynamic limit.

Furthermore, we also highlight the abnormal behavior of $\chi$ as a
function of the external driving parameter across the quantum
critical point of the Ising model and the LMG model. The possibility
of detecting QPT by QZE is raised and further exploration would be
of interest.

\section{Acknowledgement}
This work is supported by the Earmarked Grant Research from the
Research Grants Council of HKSAR, China (Project No. HKUST3/CRF/09).

\end{document}